\documentstyle[12pt]{article}

\begin{document}

\title{Epicyclic orbital oscillations in Newton's and Einstein's gravity from the geodesic deviation equation}
\vspace{0.6 cm}
\author{ {\sc Marek Biesiada} \\
{\sl Department of Astrophysics and
 Cosmology, } \\
 {\sl University of Silesia}\\
 {\sl Uniwersytecka 4,  40-007 Katowice, Poland}  \\
 mb@imp.sosnowiec.pl\\
phone: +48 32 2583 653  fax: +48 32 2660 220}

\date{}

\maketitle \vfill

\begin{abstract}
\noindent
In a recent paper Abramowicz and Klu{\'z}niak \cite{MarekWlodekGRG} have discussed the problem of epicyclic
oscillations in Newton's and Einstein's
dynamics and have shown that Newton's dynamics in a properly curved three-dimensional space is identical to test-body
dynamics in the three-dimensional optical geometry of Schwarzschild space-time. One of the main results of this paper
was the proof that different behaviour of radial epicyclic frequency and Keplerian frequency in Newtonian and
General Relativistic regimes had purely geometric origin contrary to claims that nonlinearity of Einstein's theory
was responsible for this effect.

In this paper we obtain the same result from another perspective: by representing these two
distinct problems (Newtonian and Einstein's test body motion in central gravitational field) in a uniform way
--- as a geodesic motion. The solution of geodesic deviation equation reproduces the well known results concerning
epicyclic frequencies and clearly demonstrates geometric origin of the difference between Newtonian and Einstein's
problems.

\end{abstract}


%----------------------------------------------------------------------

\section{Introduction}

In a recent paper Abramowicz and Klu{\'z}niak \cite{MarekWlodekGRG} have discussed the problem of epicyclic
oscillations in Newton's and Einstein's
dynamics and have shown that Newton's dynamics in a properly curved three-dimensional space is identical to test-body
dynamics in the three-dimensional optical geometry of Schwarzschild space-time. Their discussion was motivated
by the theory of accretion disks around black holes and neutron stars which is based on assumption that
accreting matter moves on nearly circular geodesic trajectories. One of the strong field effects that 
should be present in this context, as pointed out by Abramowicz
and Klu{\'z}niak in a recent series of papers \cite{resonance}, is the possibility of parametric resonance (preferably
3:2) between
vertical and radial epicyclic frequencies of perturbed circular orbits. It has been conjectured \cite{resonance}
that this effect is indeed responsible for the observed double peaked QPOs \cite{QPO}.

It is quite well known that in the Schwarzschild (or Kerr) spacetime radial epicyclic
frequency $\omega_r$ is lower than orbital frequency
$\omega_K$ (and vanishes at the marginally stable orbit) unlike in the Newtonian gravity where these two
frequencies are equal.
As recalled by the authors of \cite{MarekWlodekGRG} many people attributed this different behaviour to the
nonlinearity of the Einstein's theory of gravity. Therefore one of principal motivations for \cite{MarekWlodekGRG}
(besides making a brilliant use of the so called optical geometry \cite{optical})
was to demonstrate the purely geometric origin of this effect. In order to achieve this Abramowicz and Klu{\'z}niak
have represented the Einstein equations (in optical geometry) for the motion on a circular orbit in Schwarzschild
space-time in the form of Newton's equations in certain curved 3-dimensional space.
Then they were able to calculate the
epicyclic frequencies in a uniform way (i.e. from the same equation)
and show explicitly that the aforementioned difference ($\omega_r < \omega_K$ in
Einstein's gravity vs. $\omega_r = \omega_K$ in Newton's gravity) has purely geometric origin.

In this paper I will obtain the same result from another perspective: by representing these two
distinct problems (Newtonian and Einstein's test body motion in central gravitational field) in a uniform way
--- as a geodesic motion.
The difference in achieving ``uniformity'' is that whereas in \cite{MarekWlodekGRG} it was the same functional form of
the equation in our case it will be the same geometric representation of the problem.

\section{Relativistic epicycles from the geodesic deviation equation}

Before going to the details let us start with some general
comments. First of all, the problem of epicyclic frequencies has
nothing to do with nonlinearity of Einstein's equations just
because the Einstein's equations in general are dynamical
equations for evolving the 3-geometry (see e.g. \cite{MTW}). In the problem of epicyclic
oscillations around circular orbits one has a kinematic problem of
test bodies moving in static spacetime --- the geometry is static
and defined a priori. Hence the relevant question is how do the
adjacent orbits of test particles behave.

The transition from Newtonian gravity to the Einstein's picture
can be summarized in the following way.  Newton's explanation why
the planetary orbits are curved (circular, elliptical, parabolic
or hyperbolic - for comets) was that it is the force of gravity
from the central body (the Sun) that makes them curved. In
Newton's theory the nature of the force of gravity remained
unexplained - it was taken for granted.  Of course basic
properties of the gravity force were explained e.g. the inverse
square law, but not its nature. On the other hand, Einstein
attempted at explaining the nature of gravity - there is no force
field but the presence of massive central body makes the spacetime
curved. The motion of  test bodies takes place along geodesics;
they are in a free motion but in a curved spacetime, that is why
their trajectories are curved.

Therefore the (general relativistic) problem of epicyclic orbital oscillations in Schwarzschild spacetime
is exactly the problem of geodesic deviation
in Schwarzschild geometry. Stable circular orbits are stable in the sense that geodesic deviation equation solved along
such circular orbit has oscillatory solutions.

One of the most recent papers presenting solution of the geodesic deviation equation for trajectories close to circular
orbits in the Schwarzschild space-time is \cite{KernerVanHolten}. We will sketch main steps leading to the formula for
radial epicylic frequency in Schwarzschild
metric referring the interested reader to \cite{KernerVanHolten} for computational details.

In a pseudoriemannian manifold with the line element
\begin{equation}
ds^2 = g_{\mu \nu} dx^{\mu} dx^{\nu}
\end{equation}
the curve $\gamma_s := x^{\mu}(s)$ parametrized by the affine
parameter $s$ is a geodesic if the tangent vector $u^{\mu} =
\frac{d x^{\mu}}{ds}$ is paralelly transported along $\gamma_s$:
\begin{equation}
\frac{D u^{\mu}}{Ds} = \frac{d u^{\mu}}{ds} + \Gamma^{\mu}_{\nu \sigma} u^{\nu} u^{\sigma}
\end{equation}

Then consider close geodesic ${\tilde \gamma}_s$. The vector $\xi^{\mu}$ representing the separation between the
geodesic $\gamma_s$ and an adjacent geodesic ${\tilde \gamma}_s$ satisfies the geodesic deviation equation
\begin{equation} \label{geodev}
\frac{D^2 \xi^{\mu}}{Ds^2} = - R^{\mu}_{\nu \rho \sigma} u^{\nu}
\xi^{\rho} u^{\sigma}
\end{equation}
Now, let us take the Schwarzschild metric
\begin{equation} \label{Schwarzschild}
ds^2 = (1- \frac{2 G M}{c^2 r})c^2 dt^2 - \frac{1}{(1- \frac{2 G M}{c^2 r})}dr^2 -
r^2 ( d \theta^2 + \sin^2 \theta d\varphi^2)
\end{equation}
It is well known that circular orbits $r = R = const.$ are geodesics in the metric (\ref{Schwarzschild}) and test particles
move along such orbits with the angular velocity $\omega_K$ (Keplerian frequency) given by the formula \cite{MTW}
\begin{equation}
\omega_K^2 = \frac{G M}{R^3}
\end{equation}
If one considers a nearby geodesic (with respect to the circular one) and asks how does the separation between these two
behave, the answer would come from solving the geodesic deviation equation (\ref{geodev}). Technically one should express
the components of the tangent vector $u^{\mu} = (u^t,u^r,u^{\theta},u^{\varphi})$ as well as the components of the Riemann
tensor $R^{\mu}_{\nu \rho \sigma}$ (also the Christoffel symbols while calculating a covariant derivative, etc.) as
evaluated along the circular orbit.  Then one gets the system of four second-order differential equations. One of them ---
for $\xi^{\theta}$ component --- reads $\frac{d^2 \xi^{\theta}}{ds^2} = - \omega_K^2 \xi^{\theta}$ and is decoupled from
the rest. The remaining three form the system of coupled linear second order differential equations
with constant coefficients (detailed calculations can be found in \cite{KernerVanHolten})
and the characteristic equation for this system (written in matrix form) reads:
\begin{equation} \label{characteristic}
\lambda^4(\lambda^2 + \frac{G M}{R^3} (1 - \frac{6GM}{c^2 R})) = 0
\end{equation}
leading to the value of radial epicyclic oscillations with the frequency:
\begin{equation} \label{radial}
\omega_r^2 = \frac{G M}{R^3} (1 - \frac{6GM}{c^2 R})
\end{equation}
In summary, the conclusion from solving
the geodesic deviation equation in Schwarzschild spacetime is that behaviour of geodesics close to
circular orbits can be represented as a superposition of (epicyclic) oscillations around circular orbit with
two characteristic frequencies: the vertical epicyclic frequency -- equal to Keplerian frequency $\omega_K$
of the reference orbit and
radial epicyclic frequency $\omega_r$.

\section{Classical mechanics represented as problem of geodesics}

It is well known that variational principles of classical mechanics make it possible to formulate the dynamics of
Hamiltonian systems as geodesic flows on some
Riemannian manifold.
This picture comes quite naturally from the
Maupertuis-Jacobi least action principle (motion of the system with fixed energy $E$ between $q'$ and $q''$
takes place along a path $\gamma$ minimizing the Maupertuis-Jacobi action):
\begin{equation}
\delta S = \delta \;\int_{q'}^{q''} \sqrt{E-V(q)}
\sqrt{a_{ij}\;dq^idq^j} = 0 
\end{equation}
and its formal resemblance to the variational formulation of geodesics in
Riemannian geometry as curves extremalizing the distance.
This is the simplest way to see desired correspondence and it is quoted in many textbooks
on classical mechanics (e.g. in \cite{Arnold}).
Below we give some steps along a straightforward ''brute force'' derivation of this
result which could be instructive in seeing the role of time reparametrisation which is necessary in this picture.

Consider the classical mechanical system described by the
Hamiltonian
\begin{equation}  \label{OriginalHamiltonian}
H = H(p,q) = \frac{1}{2} a^{ij} p_i p_j +
V(q)
\end{equation}
The equations of
motion for the $q$ variables with respect to time
parameter $t$ (in Newtonian physics one has an absolute time) and corresponding to the
Hamilton equations
may be written as
\begin{equation} \label{Newton2law}
\ddot q^j
+ \tilde \Gamma^j_{\,ks}\,\dot q^s\,\dot q^k = - a^{ji}
 \frac{\partial V(q)}{\partial q^i}
\end{equation}
where $\tilde \Gamma^j_{\,ks}$ are the Christoffel symbols calculated with
respect to $a_{ij}$ metric and dots denote $t$ - time derivative.
Due to the force term this is, obviously, not a geodesic equation.
It is simply the Newton's second law restated.
The  momentum variables are
just linear combinations of
velocities $p_i = a_{ij} \dot q^j$.
Transformation to a geodesic motion (i.e. free motion in a curved space)
is accomplished in two steps:
$(1)$ conformal transformation of the metric $a_{ij}$, and
$(2)$ change of the time parameter along the orbit.
More explicitly we equip the configuration space
with the metric --
the so called Jacobi metric
\begin{equation}  \label{Jacobi metric}
g_{ij} = 2 (E - V(q)) a_{ij}
\end{equation}
(note that $a_{ij}$ is read off from the kinetic energy term in the
Hamiltonian and (in general) is allowed to vary as a function of
the configuration space variable, $a_{ij} = a_{ij}(q))$
). Let us also call this Riemannian space i.e. configuration space accessible for
the system and equipped with Jacobi metric
--- the Maupertuis-Jacobi manifold.\\
With respect to the metric (\ref{Jacobi metric})
and the time parameter $t$ it is not easy to see
that the orbits are geodesics since there is a
term appearing on the right hand side
of the equation,
\begin{equation} \label{notgeodesicMaupertius}
\displaystyle{d^2\over dt^2}\,q^i  +
\Gamma^i_{\,jk}\,\displaystyle{d\over dt}
q^j\,\displaystyle{d\over dt}q^k = - \frac{1}{E - V(q)}
\frac{d}{dt}q^i \frac{\partial}{\partial q^k} V(q)\frac{d}{dt} q^k
\end{equation}
where $\Gamma^i_{\,jk}$ now denote the  Christoffel symbols
associated with the Jacobi metric.
However, if we reparametrize the orbit $q^i = q^i(s)$
in terms of the parameter $s$ defined as
\begin{equation}  \label{parameterMaupertius}
d\;s =  2\;(E-V)\;d \; t
\end{equation}
the orbits will become affinely parametrized geodesics, i.e. the
configuration space variables $q^i$ satisfy the
well known geodesic equation
\begin{equation} \label{geodesicMaupertius}
\displaystyle{d^2\over ds^2}\,q^i + \Gamma^i_{\,jk}\,\displaystyle{d\over ds}
q^j\,\displaystyle{d\over ds}q^k = 0
\end{equation}
with no force term on the right hand side.

The information about the original force acting on the particle
(as described by the potential $V(q)$ in the Hamiltonian
(\ref{OriginalHamiltonian}))
has been encoded entirely  in the definition of the Jacobi metric
(\ref{Jacobi metric})
and the definition of the new parameter $s$ in
(\ref{parameterMaupertius}) parametrizing the orbit.

Contemplating how nearby orbits behave (i.e. the local instability properties), it is
natural to consider the geodesic deviation equation
which describes the behavior of nearby geodesics (\ref{geodesicMaupertius}).

This can be derived in a usual manner by subtracting the
equations for the geodesics $q^i(s)$ and $q^i(s) + \xi^i (s)$ respectively
or simply by disturbing the fiducial trajectory $(p_i(t),q^i(t))$,
\begin{eqnarray}
\tilde{p}_i(t) & = & p_i(t) + \eta_i(t), \nonumber \\
\tilde{q}^i(t) &= & q^i(t) + \xi^i(t)
\end{eqnarray}
and substituting this directly into Hamilton's equations.
In this way we also
arrive momentarily, though tediously,
at the geodesic deviation equation for the separation vector $\xi$,
\begin{equation} \label{geodesicdeviationMaupertius}
\frac{D^2\,\xi^i}{D\,s^2}=-R^i_{jkl}\,u^j \xi^k u^l
\end{equation}
Here $u^i = D q^i/ Ds $ is the tangent vector  to
 the geodesic, $\xi^j$ is the
separation vector orthogonal to $u$. 
Note, that the covariant
derivative $D / Ds $ and Christoffel symbols are calculated
with respect to the Jacobi metric (\ref{Jacobi metric}).

\section{Keplerian circular orbits represented as geodesics - epicyclic frequency in Newtonian's regime}

Geometric formulation of the Kepler problem is very simple. The two body problem in Newtonian gravity is essentially
two-dimensional.
Therefore the Jacobi metric (in polar coordinates) reads:
\begin{equation} \label{KeplerJacobi}
ds^2 = 2 (E - V(r)) (dr^2 + r^2 d\varphi^2)
\end{equation}
where: $V(r) = - \frac{G M}{r}$. 
Moreover all information carried by the Riemann curvature tensor is captured by the Gaussian curvature.

In this case the geodesic deviation equation (for an orthogonal Jacobi field $\xi$ of geodesic deviation
$g(\xi,u) = 0$) reads:
\begin{equation} \label{Gauss_deviation}
\frac{d^2}{ds^2}  {\xi}^i + K_G {\xi}^i = 0
\end{equation}
where $K_G$ is Gaussian curvature of respective Maupertuis-Jacobi manifold.

It is a simple exercise to calculate the Gaussian curvature of Maupertuis-Jacobi manifold for the Kepler problem.
Let us denote $f(r)^2 :=  2 (E - V(r))$ so that Jacobi metric reads $ds^2 = f(r)^2 (dr^2 + r^2 d\varphi^2)$, then
let us consider the one-forms
$\omega^1 := f(r)\; dr$ and $\omega^2 := f(r) r \; d\varphi$. Now, it is quite obvious that $\omega^1$ is a closed form,
and
$d \omega^2 = - \frac{1}{f(r)^2 r} \frac{d(r f(r))}{dr} \omega^2 \wedge \omega^1$. Then from Cartan's equations
one can easy read off the Gaussian curvature 
\begin{equation}
K_G = - \frac{1}{f(r)^2 r} \frac{d}{dr} \left( \frac{1}{f(r)}\frac{d(r f(r))}{dr} \right) =
 - \frac{E G M}{4(rE+GM)^3} 
\end{equation}
Note that since the kinetic energy is positive definite $T : = E - V(r) > 0$ the term $r E + GM$ is also positive and the
sign of Gaussian curvature is determined by the sign of the energy.
It is negative for $E >0$ i.e. for hyperbolic orbits and positive for $E < 0$ i.e. for bound motion.
The meaning of this results is that, in the first case, scattering of test particles on the center has sensitive dependence
on initial conditions (problem is equivalent to congruence of geodesics on negatively curved manifold) while in the
second case (equivalent to circular or elliptic orbits) the disturbed trajectories execute Keplerian epicyclic
oscillations around the orbit of reference.

For the circular orbit of radius $R$ (where $E = 1/2 V(R)$) we have:
\begin{equation} \label{Gaussian}
K_G = \frac{1}{G M R}
\end{equation}
and (by virtue of geodesic deviation equation) $K_G = \omega_{0,J} ^2$ where $\omega_{0,J}$
is the (Keplerian) epicyclic frequency
in Jacobi geometry i.e. with respect to natural ''time'' $s$ along the geodesic.

Recalling that $ds = 2(E - V(R)) dt = \frac{GM}{R} dt$ one can easily recover Keplerian epicyclic frequency in Newtonian
picture i.e. with respect to the Newtonian time $t$
\begin{equation}
\omega_0 ^2 = \frac{GM}{R^3} = \omega_K^2
\end{equation}
and it turns out to be equal to the orbital (Keplerian) frequency $\omega_K$.

Let us also remark on the equation (\ref{Gaussian}). The quantity in the denominator is in fact equal to the Keplerian
angular momentum (per unit test body mass) squared. In our approach it turns out that for circular orbits the specific
angular momentum squared is equal to the inverse Gaussian curvature of the Jacobi manifold in which the motion takes place
along geodesics. In the picture developed by Abramowicz and Klu{\'z}niak the specific angular momentum had an
interpretation of the geometric mean of the gravitational radius of central body and the radius of curvature of particle's
orbit.

\section{Conclusion}

Conclusion of this note is that by applying a uniform representation of the problem
i.e. by representing physical trajectories as a problem of geodesics in some manifold
one can see the geometric origin of the difference in epicyclic frequencies
(describing the behaviour of trajectories adjacent to circular orbits) calculated
in Newtonian and General Relativistic regimes. This result is in agreement with that of \cite{MarekWlodekGRG}
although it has been derived in a different manner. One may say that instead of applying the Feynman's principle
``the same equations have the same solutions'' we have sucessfully applied the principle of ``comparing comparable things''
by working in the same geometric representation of the problem.


\addcontentsline{toc}{section}{References}

\begin{thebibliography}{X}


\bibitem{MarekWlodekGRG}
Abramowicz M.A. and Klu{\'z}niak W., Gen. Rel. Grav. {\bf 35}, 69--77, 2003

\bibitem{resonance}
Abramowicz M.A., Almergren G.J.E., Klu{\'z}niak W., Thampan A.V., Wallinder F., Class.Quant.Grav. {\bf 19}, L52-L62, 2002\\
Abramowicz M.A., Klu{\'z}niak W., Astron. Astrophys., {\bf 374}, L19-L20, 2001\\
Klu{\'z}niak W., Abramowicz M.A., Acta Phys. Polon. B {\bf 32}, 3605-3612, 2001

\bibitem{QPO}
Strohmayer T.E., ApJ {\bf 552}, L49-L53, 2001

\bibitem{optical}
Abramowicz M.A., Carter B., Lasota J.-P., Gen. Rel. Grav. {\bf 20}, 1173, 1988

\bibitem{MTW}
Misner C.W., Thorne K.S., Wheeler J.A., {\sl Gravitation}, Freeman, New York, 1973

\bibitem{KernerVanHolten}
Kerner R., Van Holten J.W., Colistete R.,Jr., {\sl Class. Quant. Grav.}, {\bf 18}, 4725-4742, 2001

\bibitem{Arnold}
Arnold V.I., {\sl Mathematical Methods of Classical Mechanics}, Springer Verlag, 1978


\end{thebibliography}
\end{document}